**Voltage control of spin resonance in phase change materials**

Tian-Yue Chen[1], Haowen Ren[1], Nareg Ghazikhanian[2], Ralph El Hage[2], Dayne Y. Sasaki[3], Pavel Salev[4], Yayoi Takamura[3], Ivan K. Schuller[2], and Andrew D. Kent[1]

[1]Center for Quantum Phenomena, Department of Physics, New York University, New York, NY 10003, USA

[2]Department of Physics, University of California San Diego, La Jolla, CA 92093, USA

[3]Department of Materials Science and Engineering, University of California – Davis, Davis, California 95616, USA

[4]Department of Physics and Astronomy, University of Denver, Denver, Colorado 80210, USA

**Abstract**

Metal-insulator transitions (MITs) in resistive switching materials can be triggered by an electric stimulus that produces significant changes in the electrical response. When these phases have distinct magnetic characteristics, dramatic changes in spin excitations are also expected. The transition metal oxide $La_{0.7}Sr_{0.3}MnO_3$ (LSMO) is a ferromagnetic metal at low temperatures and a paramagnetic insulator above room temperature. When LSMO is in its metallic phase a critical electrical bias has been shown to lead to an MIT that results in the formation of a paramagnetic resistive barrier transverse to the applied electric field. Using spin-transfer ferromagnetic resonance spectroscopy, we show that even for electrical biases less than the critical value that triggers the MIT, there is magnetic phase separation with the spin-excitation resonances varying systematically with applied bias. Thus, applied voltages provide a means to alter spin resonance characteristics of interest for neuromorphic circuits.

**Introduction**

Voltage-induced metal-insulator transition (MIT) phenomena are being actively explored for applications in neuromorphic computing[1-5]. For example, a voltage can drive a volatile *insulator-to-metal* transition in memristive devices by forming conducting filaments parallel to the electric field and current flow, as occurs in $VO_2$[6-12] These devices have been used to produce spiking behavior in neuromorphic computing[13-16]. Materials with a *metal-to-insulator* transitions — a metallic state at low temperature and insulating state at high temperature — can show distinct behavior; an applied voltage can lead to switching into a high resistance state by the formation of a resistive barrier transverse to the applied electrical field[17-20]. Besides MIT switching devices, spintronic devices also show great potential for neuromorphic applications[21-22]. MIT switching and spintronics represent the main, yet independent, means of implementing hardware-based neurons. Integrating resistive and spintronic functionalities within a single material platform could greatly enhance the capabilities of neuromorphic circuits by harnessing the combined advantages of both effects.

The transition metal oxide $La_{0.7}Sr_{0.3}MnO_3$ (LSMO) demonstrates great potential in this regard as it is a material with a MIT[23-26] with a simultaneous magnetic phase transition[27]. The low-temperature phase of LSMO is a ferromagnetic metal with a Curie temperature of $T_c \approx 340$ K and its high-temperature phase is a paramagnet insulator[23]. Thus, a bias voltage applied to a metallic sample that drives the MIT also produces magnetic phase separation, with the formation of a paramagnetic insulating phase between ferromagnetic metallic regions[24-26]. Here we show that the spin resonance response depends strongly on the applied bias voltage and temperature. However, the response splits into multiple well-defined resonances for voltages less than the critical electrical bias ($V_c$). The spin-torque ferromagnetic resonance (ST-FMR) measurements, therefore, show that magnetic phase separation appears prior to the voltage-driven MIT. Due to the combined MIT switching and spintronic properties in multifunctional LSMO, applying electrical bias provides a sensitive new means to control spin resonance characteristics, which is of interest for spin-based neuromorphic circuits[28-30].

**Experimental Results**

**Characterization of the voltage driven MIT.** A 20 nm thick LSMO thin film was patterned into a 10 ×10 µm² device with Pd/Au electrical contacts, as illustrated schematically in Figure 1(a). Figure 1(b) shows the resistance versus temperature, which decreases with decreasing temperature below the phase change temperature of $T_c \approx 340$ K. Figure 1(c) shows the voltage-current characteristics of the device in the ferromagnetic phase at 100, 200, 250, and 300 K. At 300 K, the characteristics are nonlinear yet continuous. However, upon cooling the device below 250 K, a marked N-type negative differential resistance (NDR) can be observed. A pronounced jump and hysteretic switching at critical values could be observed at 150 K and below ($V_c = 10$ V at 150 K, $V_c = 12$ V at 100 K). The NDR indicates that the device resistance increases with increasing bias voltage, and the jump in the I-V curve corresponds to the local phase transition in LSMO producing resistive switching. Prior work on similar samples attributed this switching to localized Joule heating raising the device temperature above $T_c$[23].

**Spin-torque ferromagnetic resonance.** ST-FMR measurements were conducted on the device at 100 K using the setup illustrated in Figure 1(a), which included a DC voltage source connected via a bias tee. This configuration allows us to record the magnetic response under applied DC biases. Initially, we investigated the ST-FMR with zero DC bias applied to the sample as shown in Figure 2(a). The ST-FMR signal could be fit by a superposition of symmetric and antisymmetric Lorentzian functions[31], as shown in Figure 2(b) (for details see Supplementary S1). This analysis gives the resonance field and resonance linewidth. In Figure 2(c) the linewidths ($\Delta H$) are plotted as a function of frequency. The damping constant ($\alpha$) and inhomogeneous linewidth ($\Delta H_0$) are determined using the relation $\Delta H = \Delta H_0 + 2\pi\alpha f/\gamma$, where $f$ is the frequency and $\gamma$ is the gyromagnetic ratio[32]. The damping constant, 0.0045 and the inhomogeneous linewidth, 0.47 mT, are of the same order as those of previous reports for LSMO thin films[33-34]. Figure 2(d) shows the frequency versus resonance field. This data is fit to the Kittel model[35], $f=(\mu_0\gamma/2\pi)\sqrt{(H+H_a)(H+H_a+M_{eff})}$, to give an effective magnetization $M_{eff}$ of 0.96 T and anisotropy field $H_a$ of 13.8 mT. ST-FMR was also conducted as a function of the temperature and the inset shows how the effective magnetization depends on temperature. The decrease in effective magnetization with increasing temperature is expected for a ferromagnet approaching its Curie temperature. Overall, high effective magnetization, low damping, and narrow inhomogeneous linewidth indicate structural homogeneity and high crystalline quality of our LSMO sample.

**Voltage-dependent ST-FMR.** We used ST-FMR to determine the magnetic properties as a function of bias voltage at 100 K. The results are shown in Figure 3(a). The applied voltage varied from -15 V to +15 V to cover the range in which the voltage-driven phase transition and NDR are observed. Within the voltage range -3 to 3 V the ST-FMR spectrum exhibits a single

Lorentzian peak. The peak shifts as the voltage increases, which is attributed to Joule heating. As the applied voltage reaches ±4 V, the lineshape resembling a single peak, can no longer be fit with a Lorentzian function. Upon increasing the voltage to ±6 V, we observed that the single peak splits into three distinct peaks with the rightmost peak moving to higher field while the two lower field peaks remain nearly at fixed fields. When the applied voltage reaches the critical value of ±12 V, the higher field peak is no longer visible. The multiple peaks indicate that there are multiple magnetic phases. From previous measurements, however, phase separation is not expected to emerge at voltage below $V_c$[23]. Even above $V_c$, the expected phase separation is between the ferromagnetic matrix and paramagnetic barrier, which cannot account for the observed multiple ST-FMR peaks. We note that the same peak separation behavior was found in multiple samples (see Supplementary S2).

To analyze the data, we label the peaks in order of appearance as a function of the applied voltage L1, L2, and L3 (for details on the data analysis refer to Supplementary S3 and S4). These resonance fields are plotted in Figure 3(b). The L1 peak exhibits significant sensitivity to changes in applied voltage, whereas the L2 and L3 peaks show comparatively less change. According to the Kittel resonance condition, for fixed RF frequency, a higher resonance field indicates a lower effective magnetization. Consequently, peak L1 is indicative of parts of the device that experience heating that reduces the effective magnetization. In contrast, the L2 and L3 peaks must characterize regions that remain at a relatively lower temperature. The clear peak separation thus shows that regions with distinct magnetic properties form in response to the voltage bias. To be quantitative, we conducted measurements of the resonance field versus temperature at 6 GHz at zero voltage bias that are shown in Figure 3(c). By comparing the L1 resonance fields with the resonance field at zero bias versus temperature, we infer that resonance L1 is consistent with a region that is at 250 K when the applied bias is 11 V, while the area of L2 experiences a moderate temperature increase to 160 K. In contrast, the L3 resonance is linked to a sample region that is not significantly heated.

**Discussion**

Previous research demonstrated that in the voltage-triggered MIT of LSMO a resistive paramagnetic barrier spontaneously forms transverse to the applied electric field and expands as the voltage increases. Such phase separation was previously observed only above the critical voltage[23]. Our work, however, reveals that bias voltage can induce multiple well-defined magnetic resonances *before* the bias reaches the critical value. This result suggests distinct non-uniform magnetic properties within the LSMO sample because at equilibrium a single ferromagnetic layer with a homogeneous magnetic properties and temperature distribution exhibits only one resonance peak (see Supplementary S5).

We propose a simple model to explain the experimental observations that is illustrated schematically in Figure 4. At zero voltage, as shown in Figure 4(a), the LSMO device has a uniform temperature and magnetization distribution and thus a single ST-FMR resonance. With applied voltage, the sample heats initially without apparently inducing a significant temperature gradient or variation in magnetic properties (Figure 4(b)). The ST-FMR resonance peak shifts to higher field, indicating a heating-induced decreased magnetization. As illustrated in Figure 4(c), with increasing applied voltage there is an instability: a local region heats leading to an increase in its resistance and increased power dissipation in that region. This increased local power dissipation causes the local region to heat further, *i.e.*, this is positive feedback loop in which heating of a device region causes more heating of that region. Within this model, this positive feedback eventually leads the formation of a paramagnetic insulating barrier when the voltage bias exceeds $V_c$, as shown in Figure 4(d). As in the prior work[23], it seems reasonable to assume that the hot region is close to the middle of LSMO device, the region furthest from the high thermal conductivity electrical contacts. However, it is also possible that the phase separation starts at a structural inhomogeneity, locally more resistive sample region or sample defects.

Within this model, before reaching the critical voltage, the device has three distinct temperature zones: the hottest area, an intermediate region, and an area that remains cool, consistent with the three observed ST-FMR resonances. Upon reaching the critical voltage, the area with the highest temperature transitions into a paramagnetic state, causing its resonance peak to vanish, while the cooler areas continue to exhibit ferromagnetic properties. Therefore, only two magnetic resonances can be detected above the critical voltage. We note that this situation is distinct from typical scenarios in condensed matters physics in which materials have imperfections or are spatially inhomogeneous leading to district magnetic resonances associated with material non-uniformities. Here the formation the multiple resonances is an intrinsic characteristic associated with electric-field-induced phase separation.

## Conclusion

In conclusion, we found an unusual behavior of the ST-FMR on LSMO microstructures, indicating the presence of intrinsic voltage-induced phase separation. The single well-defined resonance at low voltages splits into multiple resonances even when the applied voltage is below the critical voltage, which induces magnetic phase separation with the formation of a paramagnetic insulating barrier. In spin oscillator-based neuron networks, the voltage control of individual oscillator frequencies is a major challenge, e.g., electrical fields only lead to small changes in the magnetic resonance characteristics of ferromagnetic transition metals. The large electrically tunable magnetic resonance in LSMO can thus provide a means to tune synapses or spin-oscillator neurons in the spintronic neural network[30]. LSMO is, therefore, a material that can be used both for the MIT switching and for spintronic applications, offering new possibilities for spintronic neuromorphic devices.

## Methods

A 20 nm thick layer of $La_{0.7}Sr_{0.3}MnO_3$ (LSMO) was epitaxially grown on a (001)-oriented $SrTiO_3$ (STO) substrate by pulsed laser deposition as described in Ref. 23. The film was patterned into 10 µm by 10 µm micron-sized features with Pd/Au electrical contacts for electrical transport and ST-FMR measurements. ST-FMR measurements were performed in the temperature range 100 to 300 K with voltage applied with a Keithley 2400 source meter. All measurements were conducted using a Quantum Design Physical Property Measurement System (PPMS). ST-FMR was conducted with an Anritsu MG3692B RF signal generator in the 4 to 12 GHz frequency range. Magnetic field was applied in plane at 45 degrees to the RF current to obtain a large signal.

## Conflict of interest

Authors declare there is no conflict of interest.

## Acknowledgements

This research was supported by the Quantum Materials for Energy Efficient Neuromorphic Computing, an Energy Frontier Research Center funded by the US Department of Energy (DOE), Office of Science, Basic Energy Sciences, under Award DE-SC0019273.

**Author contributions**

**Figures**

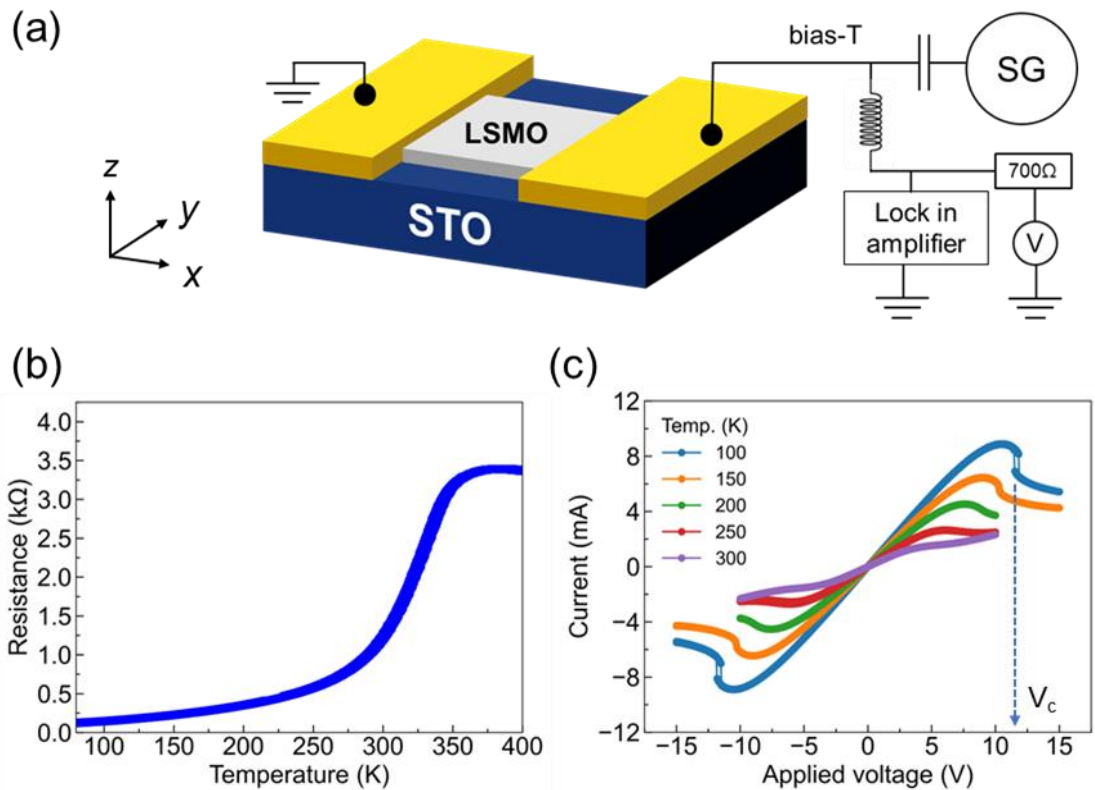

**Figure 1.** (a) Illustration of measurement setup. The LSMO device has a lateral dimension of 10 µm by 10 µm. (b) Resistance as a function of the temperature. (c) I-V curve measurements at various temperatures.

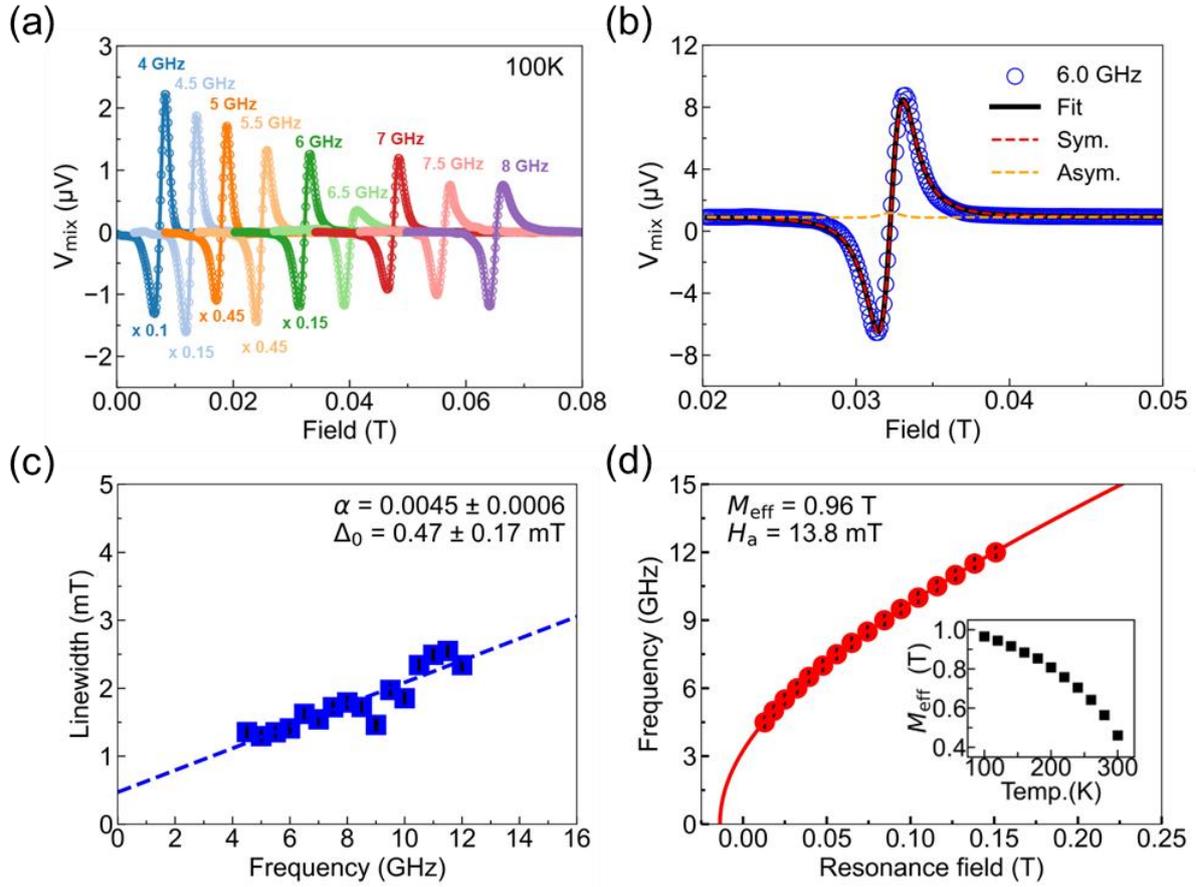

**Figure 2.** (a) ST-FMR resonance response at 100K, for frequencies from 4 to 8 GHz at zero DC bias voltage. Several curves have been scaled as indicated. (b) The 6 GHz resonance line fit to Lorentzian functions characterizing the symmetric (red) and asymmetric components (yellow) response. (c) ST-FMR linewidth as a function of frequency with a linear fit that enables determination of the damping constant α and inhomogeneous linewidth $\Delta_0$. (d) The resonant field as a function of frequency. Inset: The effective magnetization as a function of temperature.

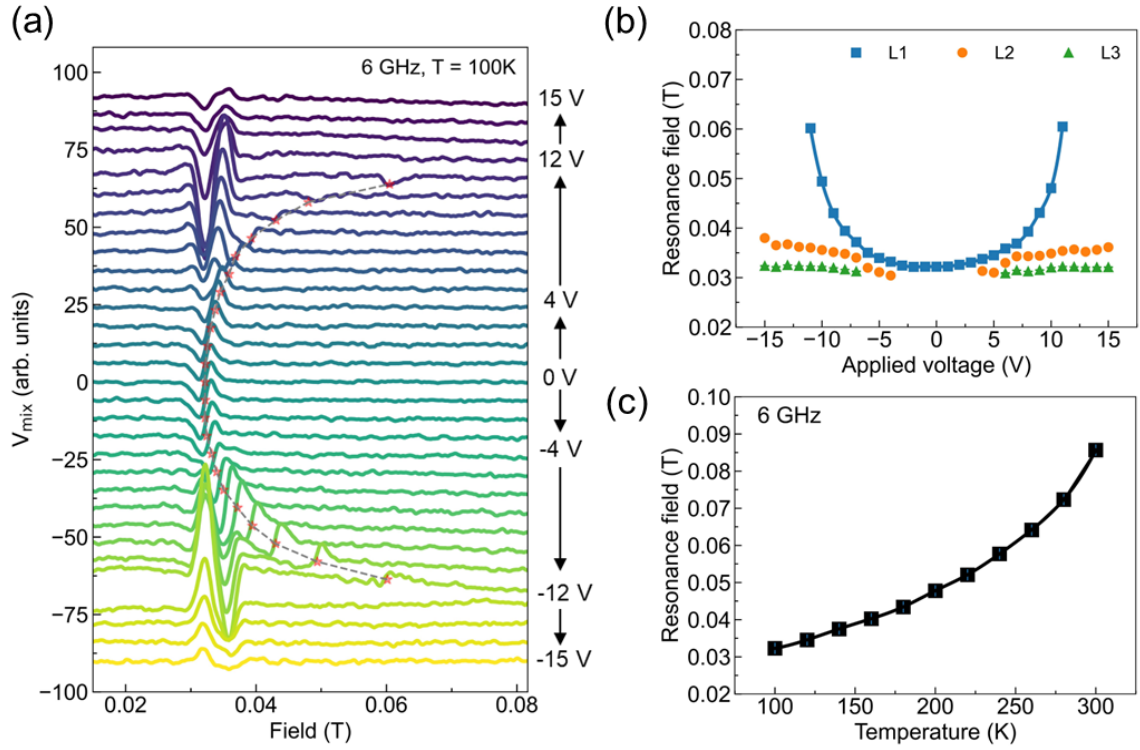

**Figure 3.** (a) ST-FMR signal as a function of bias voltage from -15 V to +15 V. (b) Summary of resonant fields L1, L2, and L3 as a function of the applied bias voltage. (c) Resonance field at 6 GHz as a function of temperature at zero bias voltage.

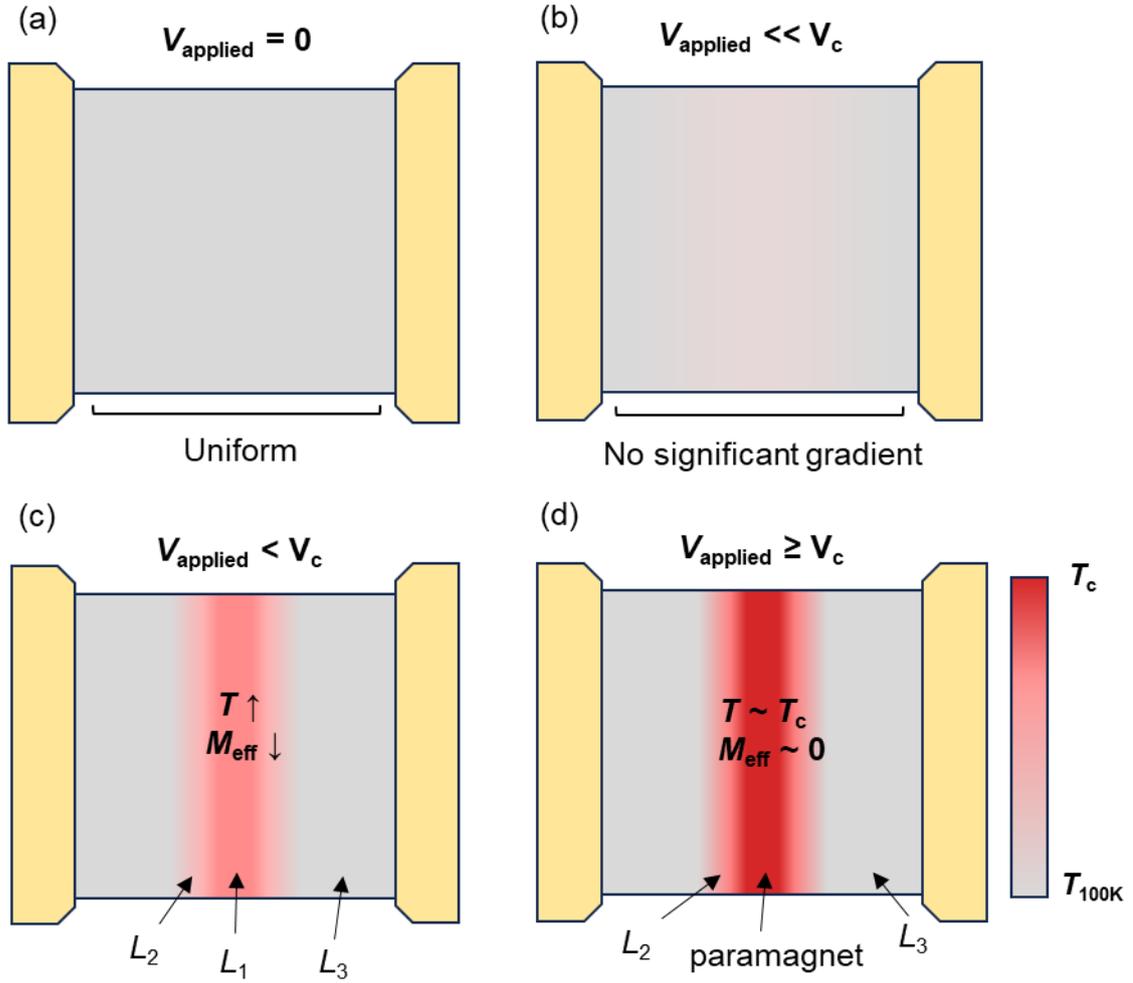

**Figure 4.** Schematic illustration of the proposed voltage-induced magnetic phase separation mechanism. (a) V = 0 V. The whole sample has a uniform temperature and effective magnetization. (b) V ≪ $V_c$. The entire sample heats but without a significant temperature gradient. (c) V < $V_c$. The center area is heated up by the applied voltage and forms a hot region with lower effective magnetization, whereas the side areas show relatively lower constant temperature. (d) V ≥ $V_c$. When the device region reaches the critical temperature, a paramagnetic resistive barrier forms where the effective magnetization drops to zero.